\documentclass[prb, twocolumn, showpacs]{revtex4}
\usepackage{amssymb}
\usepackage{amsmath}
\usepackage{graphicx}
\makeatletter

\begin{document}

\title{On the Origin of the Checkerboard Pattern in Scanning Tunneling Microscopy Maps of Underdoped Cuprate Superconductors}
 \author{Kai-Yu Yang,$^{1}$ Wei-Qiang Chen,$^{2}$ T. M. Rice,$^{1,3}$ and Fu-Chun Zhang $^{2,3}$}
\affiliation{$^{1}$ Institut f$\ddot{u}$r Theoretische Physik, ETH Z$\ddot{u}$rich,CH-8093 Z$\ddot{u}$rich, Switzerland \\
 $^{2}$ Center for Theoretical and Computational Physics and
 Department of Physics, The University of Hong Kong, Hong Kong SAR,
 China  \\
 $^{3}$   Kavli Institute for Theoretical Physics, University of California, Santa Barbara, CA 93106 USA }



\date{\today}

\begin{abstract}
The checkerboard pattern in the differential
conductance maps on underdoped cuprates appears when the scanning tunneling microscopy is
placed above the O-sites in the outermost CuO$_{\text{2}}$-plane.
In this position the interference between tunneling paths through
the apical ions above the neighboring Cu-sites leads to an
asymmetric weighting of final states in the two antinodal regions
of ${\boldsymbol{k}}$-space. The form of the asymmetry in the
differential conductance spectra in the checkerboard pattern
favors asymmetry in the localization length rather than a nematic
displacement as the underlying origin.
\end{abstract}

\maketitle

\section{Introduction}
The large scale detailed maps of the tunneling density of states
(DOS) in underdoped cuprates obtained recently by scanning
tunneling microscopy (STM) have aroused great
interest.\cite{Davis-materials-08} This surface sensitive
technique to date has been limited to strongly disordered BSCCO
and Ca$_{2-x}$Na$_{x}$CuO$_{2}$Cl$_{2}$ samples. An analysis of
these DOS-maps found a local breaking of the square symmetry of
the underlying lattice to form a checkerboard pattern. The
presence of this checkerboard pattern in these two different
underdoped cuprates has been interpreted as evidence for the
presence of an intrinsic bond-centered electronic glass with
unidirectional domains as a key characteristic of the pseudogap
phase in underdoped cuprates. \cite{Kohsaka-science-07} A number
of different interpretations of this pattern have also been
suggested in the literature. \cite{Balents-prb-05,
Tesanovic-naturephys-08, Vojta-prb-08, Granath-prb-08,
 Kim-prb-08, Pelissetto-prl-08, Vojta2-prb-08}
In our study we reexamine possible origins for this
pattern in light of very recent STM experiments. \cite{Kohsaka-nature-08}

    The new STM data taken in the superconducting state at low temperatures,
show that the checkerboard pattern appears most prominently at
voltages where the tunneling processes are predominantly into
antinodal regions in ${\boldsymbol{k}}$-space. At lower voltages the
spatial pattern of the DOS-maps is quite different. The low energy
pattern in the superconducting state was successfully analyzed as
arising from interference generated by the scattering of
propagating Bogoliubov quasiparticles in the presence of weak
disorder.\cite{Wang-prb-03} Kohsaka \textit{et al.} found that a
rapid change in the spatial pattern in the DOS-maps occurred at
the tunneling voltage corresponding to the energy of nodal
centered arcs
 of Bogoliubov quasiparticles at their endpoints on the diamond formed
 by the lines in ${\boldsymbol{k}}$-space connecting the antinodal points. \cite{Kohsaka-nature-08}
Thus the checkerboard pattern which appears at higher voltages is
formed by the tunneling of electrons and holes into pseudogap
state located in the antinodal regions of ${\boldsymbol{k}}$-space.
 The key question to be answered is the origin and interpretation of this checkerboard pattern.

      The checkerboard pattern is characterized by a local symmetry breaking
which reduces the square $C_{4}$ symmetry of the Cu-O-Cu bonds in a
(CuO)$_{4}$ square plaquette to $C_{2}$ symmetry leading to a
unidirectional pattern of domains with a glassy short range order on
the length scale of $4a$ ($a$: lattice parameter). This short range order
with local symmetry breaking has led to proposals that a
static spin-charge stripe glass coexists with the superconductivity. \cite{Vojta-prb-08, Granath-prb-08, Pelissetto-prl-08, Vojta2-prb-08}
 Another set of proposals \cite{Balents-prb-05, Tesanovic-naturephys-08, Kim-prb-08} are based on the existence of static fluctuations
in an order parameter which breaks both translational and $C_{4}$
symmetry leading to strong scattering of the Bogoliubov quasiparticles
consistent with the observations. At this point we should comment on the
difficulties associated with an explanation based on an intrinsic
symmetry breaking. While it is true that the STM experiments were
performed on the highly disordered cuprates, BSCCO and
Ca$_{2-x}$Na$_{x}$CuO$_{2}$Cl$_{2}$, because of their good surfaces.
Not all underdoped cuprates are disordered, for example, the two members
of the YBCO family, YBa$_{2}$Cu$_{4}$O$_{8}$ and YBa$_{2}$Cu$_{3}$O$_{6.5}$-ortho-II,
are well ordered and intrinsically strongly underdoped with hole concentrations estimated to be
$x = 0.14$  and  $0.1$ respectively. Indeed samples of these cuprates
 are of sufficient quality to allow the observation of quantum oscillations
 at high fields and low temperatures. \cite{Leboeuf-nature-07}
Zero field experiments on these well ordered samples show no
evidence of static broken translational symmetry in nuclear
magnetic resonance (YBa$_{2}$Cu$_{4}$O$_{8}$) \cite{Tomeno-prb-94}
or neutron scattering (YBa$_{2}$Cu$_{3}$O$_{6.5}$-ortho-II )
experiments. \cite{Stock-prb-02} This lead us to conclude that the
translational symmetry breaking observed by STM in the BSCCO and
Ca$_{2-x}$Na$_{x}$CuO$_{2}$Cl$_{2}$ samples results from the
strong disorder associated with random doping a short range spin liquid
in these cuprates.

  In this paper we propose that the checkerboard pattern observed by STM
is caused by the local disorder in the CuO$_{2}$ planes. Our
explanation goes back to the observation by Chen, Rice and Zhang
(CRZ) \cite{chen-prl-06} that when the STM tip is centered above
the planar O-sites, it couples to two tunneling paths through
apical Cl ions above the neighboring Cu-sites and as a consequence
interference between these paths plays an important role. This
interference leads to a different tunneling matrix elements for
electrons into hole states which are bonding (anti-bonding) upon
reflection about the central O-site of the Cu-O-Cu bond.
\cite{chen-prl-06} CRZ considered the dilute limit and examined
the localized states of a single hole bound to a Na$^{+}$-acceptor
in a Ca$_{2-x}$Na$_{x }$CuO$_{2}$Cl$_{2}$ sample. The
Na$^{+}$-acceptor sits above the center of a (CuO)$_{4}$ square
plaquette. They found a doubly degenerate acceptor bound state
which in the presence of a local quadrupole electric field splits
into two states, each with only $C_{2}$ bonding symmetry. However
this explanation of the local symmetry reduction from $C_{4}$ to
$C_{2}$ symmetry is not compatible with the recent STM data, which
shows that the checkerboard pattern is formed by the tunneling of
electrons and holes into pseudogap states located in the antinodal
regions of ${\boldsymbol{k}}$-space. The lowest energy for a
single hole is at the nodal regions of
${\boldsymbol{k}}$-space so that single hole acceptor bound
states are made up from combinations of hole states near to the
nodal not antinodal regions of ${\boldsymbol{k}}$-space.
Nonetheless their main conclusion that interference of tunneling
paths occurs when the STM tip is above an O-site, remains valid.

\begin{figure}[tbp]
\centering
\begin{minipage}[t]{1.0\linewidth}
\centering
\includegraphics[width = 6cm, height = 5cm, angle= 0]{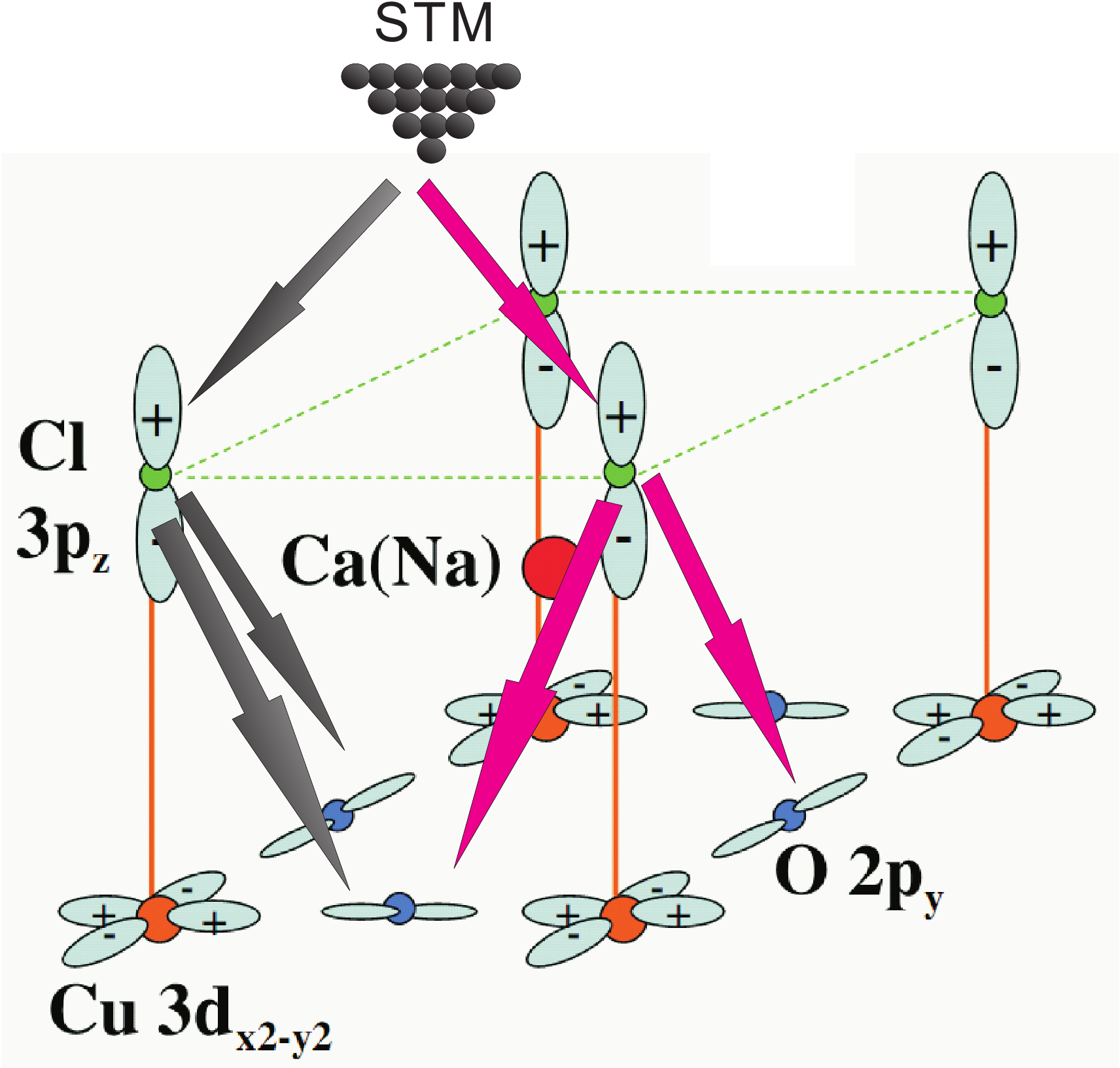}
\end{minipage}%

\begin{minipage}[b]{1.0\linewidth}
\centering
\includegraphics[width=7.5cm,height=5.5cm, angle=0]
{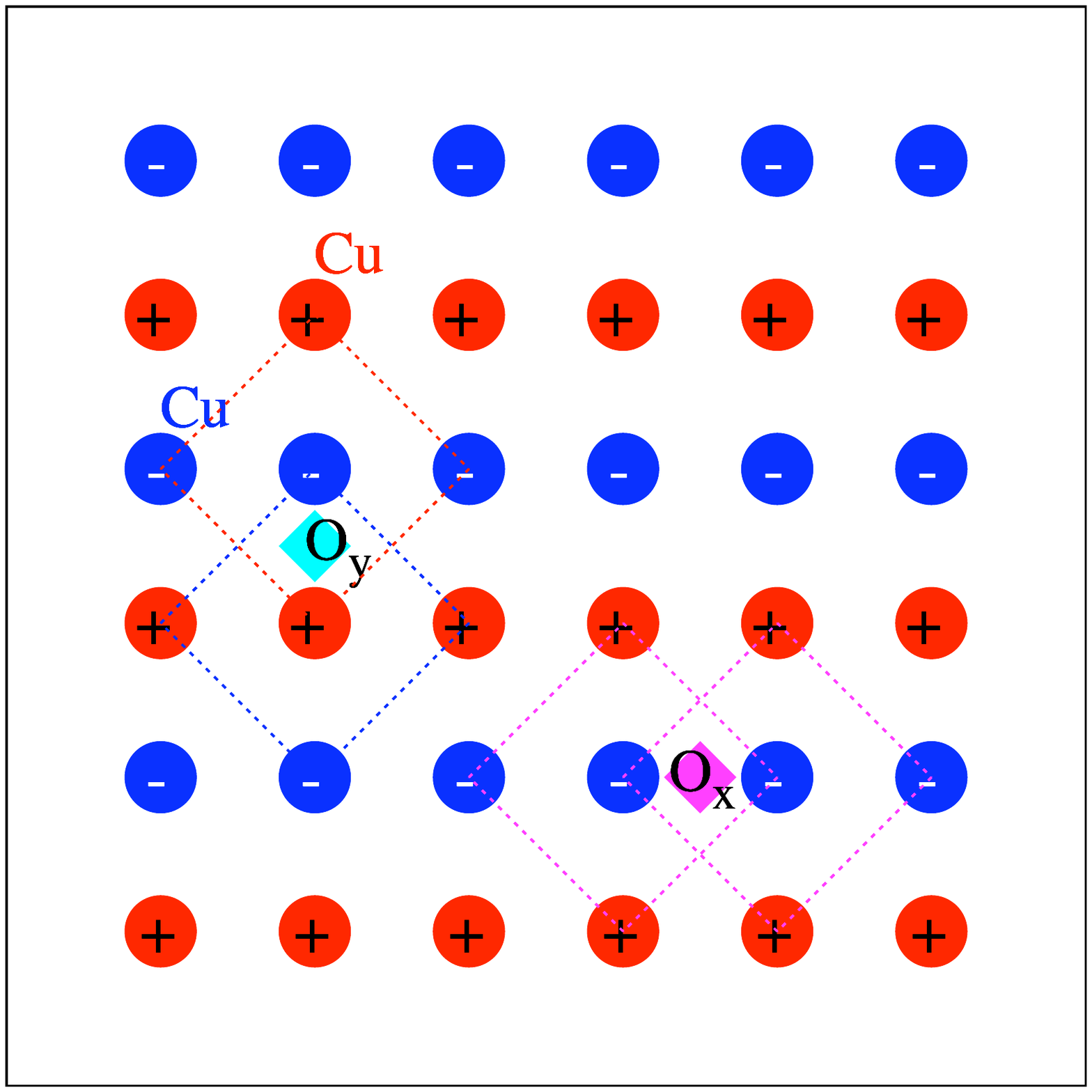}
\end{minipage}

\caption{(Color online) Top panel: schematic demonstration of the
interference of the possible two tunneling paths from STM tip to
the Cu-O-Cu hybrid state when the tip is at the midway between two
Cl atoms. Lower panel: the contribution of the antinodal
${\boldsymbol{k}}_{A,y}=(0,\pi)$ state to the two interference
paths. The blue and red dots are Cu sites with the alternative
sign along $\boldsymbol{y}$-direction denoting the sign of the Bloch wave
function of antinodal ${\boldsymbol{k}}_{A,y}$ quasiparticles
as shown in Eq.\ref{eq:anticreator}. (The signs for the
$d_{x^{2}-y^{2}}$ symmetry of Cu-orbitals are not shown.) For the
${\boldsymbol{k}}_{A,y/x}$ quasiparticle contributions to
O$_{\text{y/x}}$ of the two paths cancel each other, meanwhile the
contributions to O$_{\text{x/y}}$ of the two paths add together. }

\label{fig:stm_path}
\end{figure}

 In light of this new STM data \cite{Kohsaka-nature-08}
we concentrate our attention on the
role of the antinodal regions of ${\boldsymbol{k}}$-space as origins of
the checkerboard pattern. There are two such regions, one near ($\pi$,0)
and the other near (0,$\pi$). The bonding-antibonding pattern on the
Cu-O-Cu bonds differs between the two antinodal regions.
For quasiparticle near the (0,$\pi$) antinodal region the Bloch phase
 factor is bonding on Cu-O-Cu in the $\boldsymbol{x}$-direction and anti-bonding in
the $\boldsymbol{y}$-direction, as shown in Fig.\ref{fig:stm_path}(b). This pattern is reversed for the case of the
($\pi$,0) antinodal region. Thus when the STM tip lies above the
planar O-sites in a (CuO)$_{4}$ square plaquette the stronger
signal comes from the antinodal region with a bonding phase factor
and a local difference in the tunneling DOS between the two
antinodal regions will
 cause a breaking of the $C_{4}$  to $C_{2}$ symmetry. This symmetry
breaking occurs only when the STM tip lies above the O-sites in a
(CuO)$_{4}$ square plaquette but not when the tip is above Cu-sites.
Typical checkerboard patterns display strong modulation above the
planar O-sites but much weaker modulation is observed at the Cu sites.
This strongly suggests that we should concentrate on possible mechanisms
that locally differentiate the two antinodal regions since it is this
difference which will show up at the planar O-sites.

In their recent paper, Kohsaka \textit{et al.} discussed the
conditions that enhance the checkerboard pattern in the tunneling
maps. They find that the pattern emerges most clearly in $Z$-maps
which plot the ratio of differential conductances $g({\boldsymbol{r}},V)$, at
opposite bias, \textit{i.e.}
\begin{eqnarray}
Z({\boldsymbol{r}}, E=eV) &=& g({\boldsymbol{r}},+V) / g({\boldsymbol{r}},-V)     \label{eq:Zmap}
\end{eqnarray}
Further the checkerboard pattern is most pronounced when the voltage $V$
is adjusted locally to the local pseudogap value.
This value varies strongly across the field of view in the STM measurements.
This is typically $40a \times 40a$ and the typical length scale of the
variation in the local value of the pseudogap is roughly $4a$,
similar to the length scale of coherent checkerboard patterns.

In this paper we examine possible sources of local symmetry
breaking from $C_{4}$ to $C_{2}$ symmetry in the STM patterns in the
presence of strong local disorder, combining the CRZ theory for
STM tunneling with the Yang, Rice and Zhang (YRZ)
\cite{Yang-prb-06} phenomenological form for the single electron
propagator in the underdoped pseudogap phase.

\section{YRZ-Phenomenological Single Particle Propagator}
YRZ based their propagator on a generalization of the form of the propagator
for a lightly doped array of 2-leg Hubbard ladders derived by Konik
\textit{et al.} \cite{Konik-prl-06}. They introduced a self-energy,
$\Sigma_{R}({\boldsymbol{k}}, \omega, x)$, which diverges at $\omega=0$ on a
surface spanned by elastic particle-particle umklapp scattering
analogous to the behavior of the ladder model.
In the 2D square lattice this umklapp surface is a diamond
connecting antinodal points on the Brillouin zone boundary.
Note this umklapp surface appears as the energy gap surface also
in the case of wider Hubbard ladders with more than 2 legs \cite{Lehur-arxiv-08}.
YRZ took over the form of $d$-wave RVB gap function
$\Delta_{R}({\boldsymbol{k}},x)=\Delta_{0}(x) (\cos{k_{x}} -\cos{k_{y}})$
from the renormalized mean field theory of Zhang \textit{et al.} \cite{Zhang-88}
Using their Gutzwiller renormalization factor $g_{t}(x)$
for longer range hopping, YRZ proposed a form for the coherent part of the single particle propagator  $G^{P}({\boldsymbol{k}},\omega, x)$
in the pseudogap state,
\begin{eqnarray}
G^{P}({\boldsymbol{k}},\omega,x) &=& \frac{g_{t}(x)} {\omega - \epsilon({\boldsymbol{k}}) -\mu -\Sigma_{R}({\boldsymbol{k}},\omega,x)}  \notag \\
&=& g_{t}(x) \sum_{i} \frac{Z^{i}_{{\boldsymbol{k}}}} {\omega - E^{i}_{{\boldsymbol{k}}}}\notag \\
\Sigma_{R} {({\boldsymbol{k}},\omega,x)} &=& \frac{\Delta_{R}^{2}({\boldsymbol{k}},x)}
{\omega + \epsilon^{0}({\boldsymbol{k}})}  \label{eq:YRZ}
\end{eqnarray}
where $\epsilon^{0}({{\boldsymbol{k}}})$ only includes the NN renormalized hopping contribution,
meanwhile $\epsilon({{\boldsymbol{k}}})$ includes both the NN and longer range renormalized hopping
contributions.

The YRZ propagator fits well the quasiparticle dispersion $E^{i}_{\boldsymbol{k}}$ and their weight $Z^{i}_{\boldsymbol{k}}$ seen in angle
resolved photoemission (ARPES) experiments on the pseudogap phase. \cite{Yang-prb-06, Yang-epl-09} It has aslo been used successfully to interpret a number of other experiments on the cuprate superconductor, \textit{i.e.} Raman scattering \cite{Bascones-prl-07}, optical properties \cite{Illes-prb-09}, and specific heat \cite{Leblanc-prb-09}. 
Recently Yang \textit{et al.} \cite{Yang-epl-09} showed that it also accounts well
for the particle-hole asymmetry in the quasiparticle properties opening up along the Fermi arcs reported
by Yang \textit{et al.} in the normal pseudogap phase. \cite{Yang-nature-08} 
Very recently \cite{Carbotte-arxiv-09} the temperature and doping dependence of superconducting penetration depth was also shown to fit well to this scheme. 
In the antinodal regions of  ${\boldsymbol{k}}$-space the energy gap is
largest with a magnitude
 $\Delta_{R}({\boldsymbol{k}}_{A},x )$, where ${\boldsymbol{k}}_{A}$
is an antinodal wavevector: ${\boldsymbol{k}}_{A,x} = (\pi,0)$ and
${\boldsymbol{k}}_{A,y}=(0,\pi)$.  The value of $\Delta_{0}(x)$ is taken from the renormalized mean
field theory of Zhang \textit{et al.} for the RVB state. \cite{Zhang-88}
It is a strong function of the hole density, $x$, dropping linearly with
increasing $x$ from  $\Delta_{0}$ at $x=0$ to 0
 at a critical hole concentration $x=x_{c} (\sim 0.2)$.

\section{Random Electric Field in a Thomas-Fermi Approximation}
\begin{figure}[tbp]
\centering
\begin{minipage}[b]{0.8\linewidth}
\centering
\includegraphics[width=7cm,height=4.5cm, angle=0]
{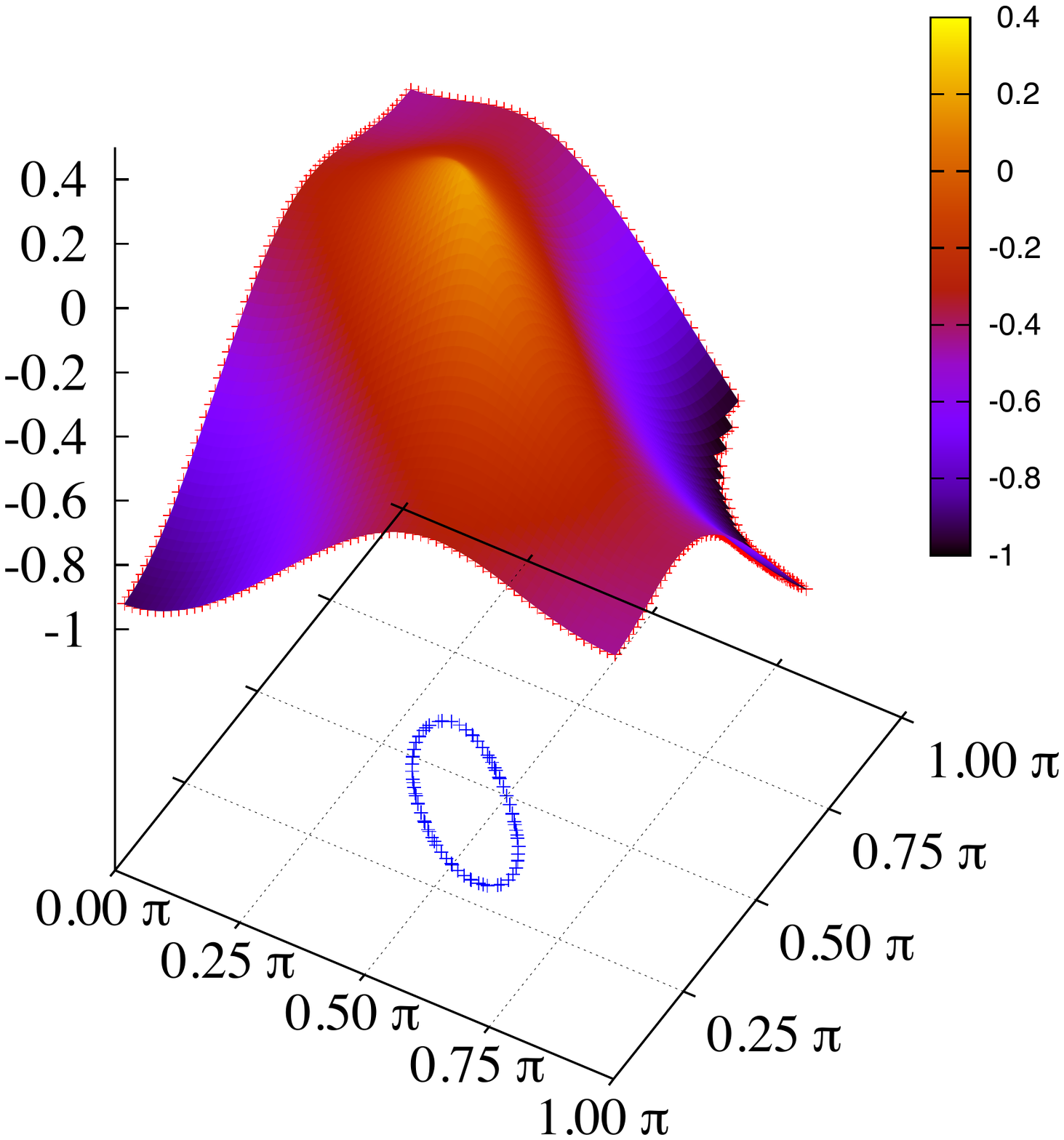}
\end{minipage}

\begin{minipage}[t]{0.8\linewidth}
\centering
\includegraphics[width=7cm,height=3cm, angle=0]
{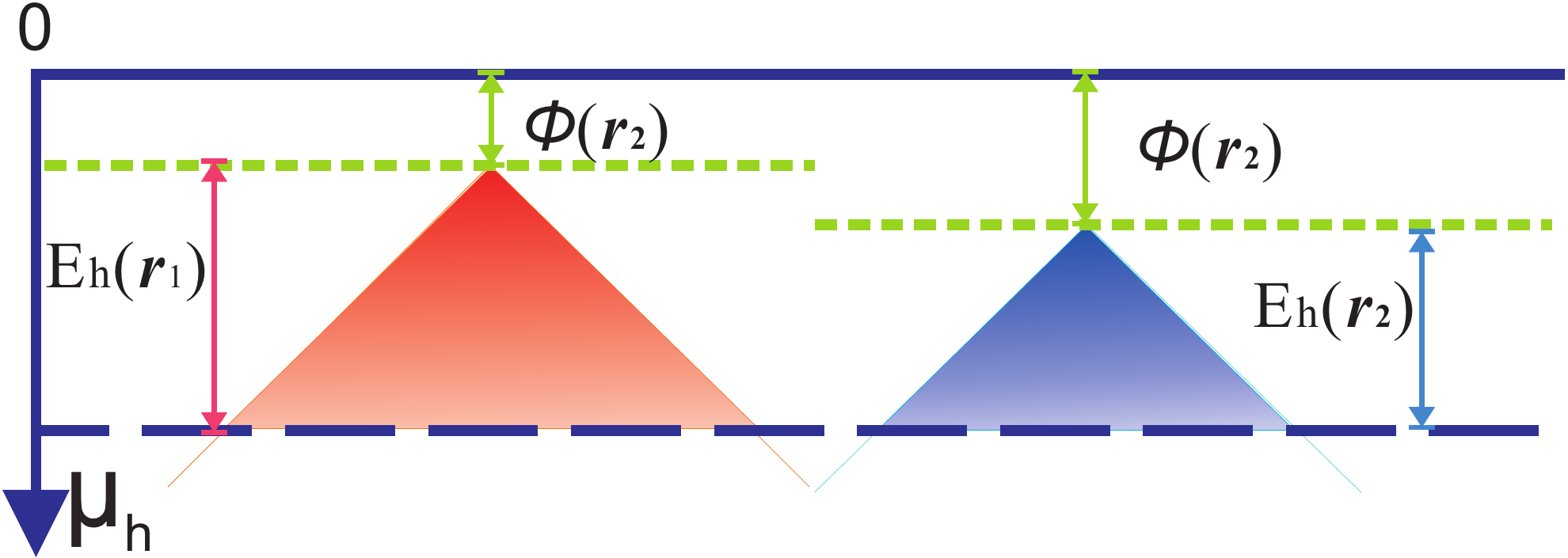}
\end{minipage}%
\caption{(Color online)  Top Panel: the low band from the single
particle propagator Eq.\ref{eq:YRZ} in pseudogap state with the
Fermi surface (hole pocket) projected on the xy plane. The bottom
of the hole fermi sea is located at the midway between the node
and $(\pi/2, \pi/2)$. Lower panel: schematic demonstration of the
spatial variation of the hole Fermi energy $E_{h}
({\boldsymbol{r}})$ due to the random electric field  $\phi({\boldsymbol{r}})$. }
\label{fig:thomas-fermi}
\end{figure}

In the cuprates BSCCO and Ca$_{2-x}$Na$_{x}$CuO$_{2}$Cl$_{2}$
which have been studied in the STM experiments, the holes are
introduced by randomly distributed acceptors situated above and
below the CuO$_{2}$-planes. The random electric field  $\phi({\boldsymbol{r}})$
associated with the acceptors generates a random component in the
local hole density. The case of a slowly varying potential can be
treated in a Thomas-Fermi approximation. The hole chemical potential
$\mu_{h}$ is a constant in space. In the YRZ propagator the hole
pocket is determined by the quasiparticle energy contour at the
chemical potential $\mu_{h}$
which lies above the minimum hole
energy $E^{b}_{h}$ located at the midway between the node
and $( \pi/2, \pi /2$) points in hole notation as shown in the upper panel of Fig. \ref{fig:thomas-fermi}.
The hole density of states (DOS) rises almost linearly with increasing hole
energy in the YRZ ansatz, in agreement with recent angle
integrated photoemission data. \cite{Hashimoto-arxiv-08}
The spatial varying hole Fermi energy $E_{h}({\boldsymbol{r}})$ relative to the bottom of the hole band
in a simple Thomas-Fermi ansatz is given by
\begin{eqnarray}
E_{h}({\boldsymbol{r}})&=& \mu_{h} - \phi({\boldsymbol{r}}) \label{eq:TF}
\end{eqnarray}
leading to a spatial variation in the hole density
$x{({\boldsymbol{r}})}$, as shown in the lower panel of Fig. \ref{fig:thomas-fermi}. This has two effects,
first, a breathing of the area enclosed by the hole pockets
and therefore the Fermi arcs observed by ARPES and by quasiparticle
interference in STM.
Second, it introduces a local variation in the RVB gap
$\Delta_{R}({\boldsymbol{k}}_{A}, x)$ which as remarked above,
is a strong function of the hole density.
The strong spatial variation in the RVB gap can be measured
by the energy of the maximum in the DOS
relative to the constant chemical potential.
That has been reported in several STM studies. \cite{Kohsaka-science-07, Kohsaka-nature-08, Wise-naturephys-09} A breathing of the Fermi arc determined by
quasiparticle interference in STM spectra, was recently reported
by Wise \textit{et al.} \cite{Wise-naturephys-09}


\section{ Interference in Tunneling Spectra with the STM Tip above O-sites}
The detailed STM maps showing the checkerboard patterns were taken
on two compounds, Dy-doped BSCCO and
Ca$_{2-x}$Na$_{x}$CuO$_{2}$Cl$_{2}$. In BSCCO samples the outmost surface layer is a BiO plane with the Bi-ions located directly above the Cu-sites in the CuO$_{2}$-plane. Assuming the STM tips couple predominantly to the empty $6s$/$6p$ Bi-orbitals, the analysis presented below  will apply also to BSCCO. We shall restrict ourselves
to the former which has a simpler structure. At this point it is
useful to recap the CRZ analysis, which discussed the role
interference between neighboring tunneling paths in the STM maps. Meanwhile, we focus on the antinodal quasiparticles.
With the STM tip centered at position ${\boldsymbol{r}}$, the
differential conductivity at T=0 and voltage $V$, is following
Tersoff and Hamann, \cite{Tersoff-prb-85}
\begin{eqnarray}
\frac{dI({\boldsymbol{r}})} {dV} &\propto& \sum_{\sigma, m} \mid
\langle m \mid a^{\dag}_{{\boldsymbol{r}},\sigma} \mid \Psi _{0} \rangle \mid^{2}
\delta(\omega -E_{m}+E _{0}) \label{eq:stm}
\end{eqnarray}
where  $a^{\dag}_{{\boldsymbol{r}},\sigma}$ is the electron creation operator
at position ${\boldsymbol{r}}$, $\mid m \rangle$ are final eigenstates
with energy $E_{m}$, and $\omega = eV$, $\mid \Psi _{0} \rangle$
is the ground state with energy $E _{0}$. The outermost surface layer
is composed of Cl-ions which sit directly above the Cu-ions in the
topmost CuO$_{2}$-layer, so that the STM tip couples primarily
to their $3p_{z}$-states (see Fig 1).
When the STM tip is scanned from above the Cl-ion at ${\boldsymbol{i}}$ to a NN Cl-ion site,
${\boldsymbol{i}}+\boldsymbol{\tau}$ [$\boldsymbol{\tau}=(\pm 1, 0)$ or $(0, \pm 1)$], we have
\begin{eqnarray}
a^{\dag}_{{\boldsymbol{r}},\sigma}  &=&
 \langle {\boldsymbol{i}}, \text{Cl} \mid {\boldsymbol{r}}  \rangle p^{\dag}_{\text{Cl}, {\boldsymbol{i}},\sigma}
 + \langle  {\boldsymbol{i}}+\boldsymbol{\tau}, \text{Cl} \mid {\boldsymbol{r}} \rangle p^{\dag}_{\text{Cl}, {\boldsymbol{i}}+\boldsymbol{\tau},\sigma}  \notag \\
p^{\dag}_{\text{Cl}, {\boldsymbol{i}},\sigma} &=&   \sum_{\boldsymbol{\tau}^{\prime}} \langle {\boldsymbol{i}}+\boldsymbol{\tau}^{\prime}, \text{Cu} \mid {\boldsymbol{i}}, \text{Cl} \rangle  c^{\dag}_{{\boldsymbol{i}}+\boldsymbol{\tau}^{\prime}, \sigma} \notag \\
& \sim &\frac{1}{2} \sum_{\boldsymbol{\tau}^{\prime}} (-1)^{{\tau^{\prime}_{x}}}   c^{\dag}_{{\boldsymbol{i}}+\boldsymbol{\tau}^{\prime}, \sigma}
\label{eq:interfernce}
\end{eqnarray}
where $(-1)^{{\tau^{\prime}}_{x}}$ is from the $d_{x^{2}-y^{2}}$ symmetry of the Cu-orbital,
$p^{\dag}_{\text{Cl}, {\boldsymbol{i}},\sigma}$ is the electron creation operator
at site ${\boldsymbol{i}}$ Cl, and $c^{\dag}_{{\boldsymbol{i}}, \sigma}$ is the electron creation operator
for the $d$-$p$ hybridized orbital centered on the Cu-site at
${\boldsymbol{i}}$. The integrated current up to a positive voltage $V$
(electron injection) is then
\begin{eqnarray}
I({\boldsymbol{r}},\omega) &=& \sum_{\sigma,m} \mid \langle m \mid \sum_{\boldsymbol{\tau}^{\prime}}(-1)^{\tau^{\prime}_{x}}
\big[ \langle {\boldsymbol{i}},\text{Cl} \mid  {\boldsymbol{r}} \rangle c^{\dag}_{{\boldsymbol{i}}+\boldsymbol{\tau}^{\prime}, \sigma} \notag \\
&& + \langle {\boldsymbol{i}}+\boldsymbol{\tau}, \text{Cl} \mid  {\boldsymbol{r}}  \rangle c^{\dag}_{{\boldsymbol{i}}+\boldsymbol{\tau}^{\prime}+\boldsymbol{\tau}} \big] \mid \Psi _{0} \rangle \mid^{2} \notag \\
&& \Theta(\omega-E_{m}+E _{0}) \label{eq:current}
\end{eqnarray}
$\Theta$ is the Heaviside step function. CRZ pointed
out that, when the tip is positioned above the O-ion at ${\boldsymbol{i}} + \boldsymbol{\tau}/2$ lying halfway
between the two Cu-ions at ${\boldsymbol{i}}$ and ${\boldsymbol{i}}+\boldsymbol{\tau}$, the integrated current will be sensitive
to the relative phase of the hole states centered at the NN Cu
-ions.  Note the orthogonality of the Cl $3p_{z}$-states to the
$d_{x^2-y^2}$-symmetry of the hole states centered on the Cu-site underneath
leads to a dominant hybridization with the hole states centered on
the 4 NN Cu-sites. \cite{Balatsky-rmp-06} Finally, we arrive at the result for the
tunneling currents when the STM tip lies halfway between NN Cl sites, i.e. above the O-sites in the
$\boldsymbol{x}/\boldsymbol{y}$ oriented  Cu-O-Cu bonds $I^{O_{x/y}}({\boldsymbol{r}}+\boldsymbol{\tau}/2, \omega)$ and $I^{Cu}({\boldsymbol{r}}, \omega)$
when the STM tip lies above a Cu site
\begin{eqnarray}
I^{O_{x/y}}
&=& \Theta(\omega-E_{m}+E _{0}) \notag \\
&& \mid \langle m \mid \sum_{\boldsymbol{\tau}^{\prime}, \sigma}(-1)^{\tau^{\prime}_{x}} \big(c^{\dag}_{{\boldsymbol{i}}+\boldsymbol{\tau}^{\prime}, \sigma}
+ c^{\dag}_{{\boldsymbol{i}}+\boldsymbol{\tau}+\boldsymbol{\tau}^{\prime}} \big) \mid \Psi _{0} \rangle \mid^{2}  \notag  \\
I^{Cu}
&=&  \Theta(\omega-E_{m}+E _{0}) \notag \\
&&\mid \langle m \mid \sum_{\boldsymbol{\tau}^{\prime},\sigma} (-1)^{\tau^{\prime}_{x}} c^{\dag}_{{\boldsymbol{i}}+\boldsymbol{\tau}^{\prime}, \sigma}
 \mid \Psi _{0} \rangle \mid^{2}  \label{eq:Cucurrent}
\end{eqnarray}
 CRZ considered the case of a single hole bound to a Na$^{+}$-acceptor which they showed may well have a
  doubly degenerate ground state. This degeneracy in turn splits in the presence of a quadrupole electric field,
  into states with reflection symmetries with respect to the $x$ and $y$ axes $P_{x} = \pm 1$, $P_{y} = \pm 1$ \textit{i.e.} bonding
  and anti-bonding states at the O-sites. They showed that this lower $C_{2}$ symmetry shows up in the tunneling current.

 As we remarked earlier, the new STM data on underdoped Ca$_{2-x}$Na$_{x}$CuO$_{2}$Cl$_{2}$ samples show that the checkerboard pattern
 is associated into antinodal states rather than the single hole bound states considered by CRZ. Note, however, quasiparticles in the antinodal
 regions states are composed of Bloch states near  ${\boldsymbol{k}}_{A,x} = (\pi,0)$ and  ${\boldsymbol{k}}_{A,y} = (0,\pi)$, and these have similar
 reflection properties with opposite parities for reflections about the $\boldsymbol{x}$- and $\boldsymbol{y}$- axes. Focusing on the contributions from
 antinodal quasiparticles, we can set
\begin{eqnarray}
c^{\dag}_{{\boldsymbol{i}}, \sigma} &\sim& c^{\dag}_{{\boldsymbol{k}}_{A,x}} (-1)^{i_{x}}
+  c^{\dag}_{{\boldsymbol{k}}_{A,y}} (-1)^{i_{y}}   \label{eq:anticreator}
\end{eqnarray}
with the sign alternating pattern from ${\boldsymbol{k}}_{A,y}$ shown in Fig.\ref{fig:stm_path}(b). So that when the tip is above the
O$_{\text{x/y}}$ site at ${\boldsymbol{i}} +\boldsymbol{\tau}/2$
\begin{eqnarray}
a^{\dag}_{{\boldsymbol{i}}+\boldsymbol{\tau}/2,\sigma}
&\sim&  \frac{1}{2} \sum_{\boldsymbol{\tau}^{\prime}} (-1)^{\tau^{\prime}_{x}} \big[ c^{\dag}_{{\boldsymbol{i}}+\boldsymbol{\tau}^{\prime}, \sigma}
 + c^{\dag}_{{\boldsymbol{i}}+\boldsymbol{\tau}^{\prime}+\boldsymbol{\tau}, \sigma} \big]  \notag \\
&\sim&  \sum_{{\boldsymbol{k}}_{A, x/y}, \boldsymbol{\tau}^{\prime}}  (-1)^{\tau^{\prime}_{x}}  c^{\dag}_{{\boldsymbol{k}}_{A,x/y}}
(-1)^{({\boldsymbol{i}}+\boldsymbol{\tau}^{\prime})_{x/y}} \big[ 1+(-1)^{\tau_{x/y}} \big] \notag \\ \label{eq:antiQP}
\end{eqnarray}
From the factor $1+(-1)^{\tau_{x/y}}$, it is obvious that for
$O_{x/y}$ the contribution from the ${\boldsymbol{k}}_{A,x/y}$
quasiparticles vanishes, meanwhile that from
${\boldsymbol{k}}_{A,y/x}$ is enhanced due to the interference
effect, as schematically shown in Fig.\ref{fig:stm_path} As a
result any local perturbation which breaks
 the symmetry between the two antinodal regions of ${\boldsymbol{k}}$-space will lead to a lower $C_{2}$ symmetry in the tunneling current.
 This leads us then to look at various sources that can give rise to such local symmetry breaking in this highly random system. To proceed
 further it is useful to examine the nature of the anisotropy observed in the tunneling spectra more closely.

\section{Measured Anisotropic STM Tunneling Spectra.}
\begin{figure}[t]
\centerline{\includegraphics[width = 6.5cm, height = 6.5cm, angle
= 0]{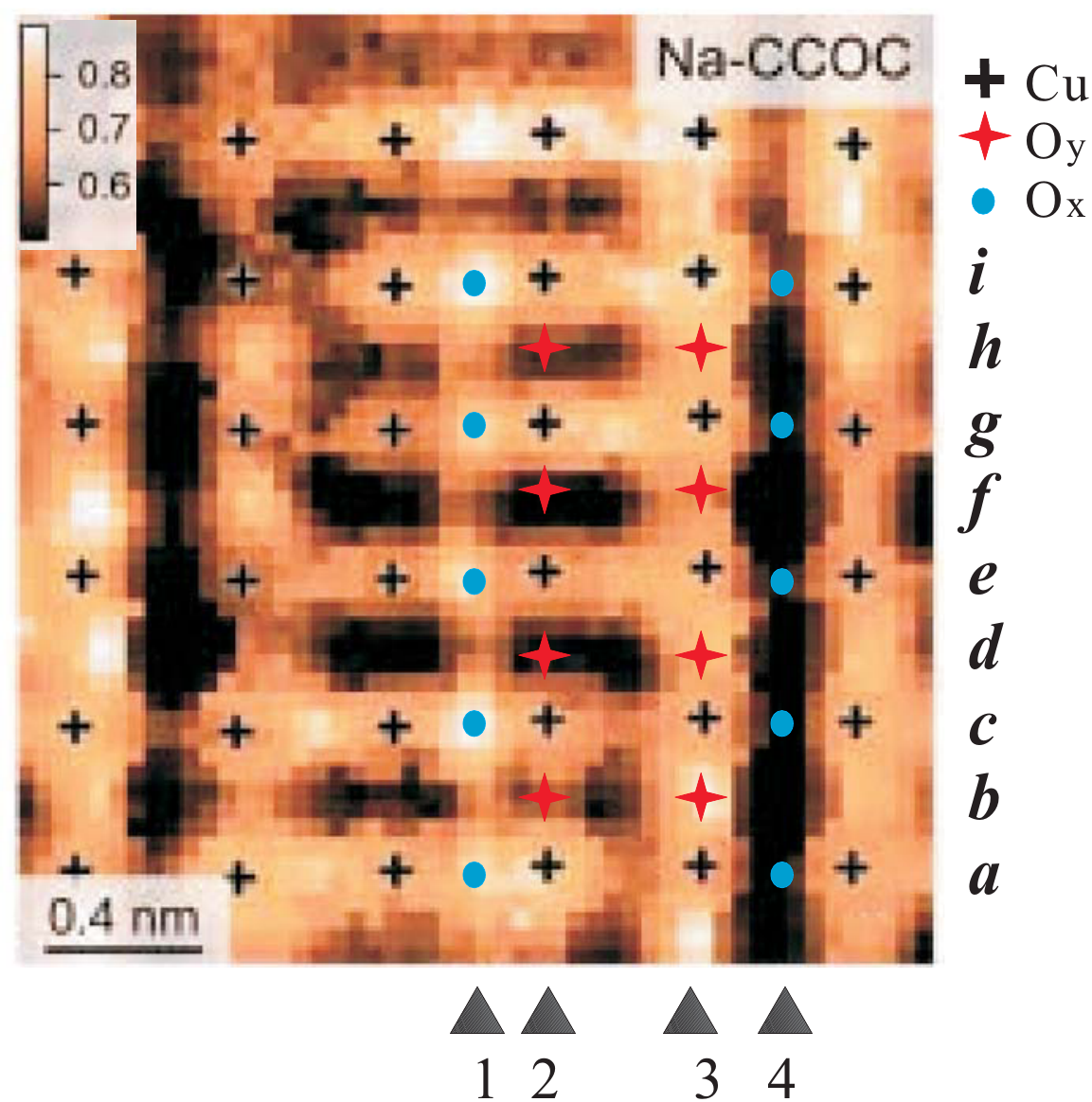}} \caption[]{ (Color
online) Typical checkerboard pattern revealed by R-map in Na-CCOC
observed in one recent STM experiments
Ref.\cite{Kohsaka-science-07} (reproduced from original paper).
The blue solid circles (red solid stars) show the position of the
oxygen atoms on Cu-O-Cu bond along $\boldsymbol{x}/\boldsymbol{y}$ directions [labelled by
the combination of vertical lines (1-4) and horizon (a-i)].}
\label{fig:STM}
\end{figure}

Kohsaka \textit{et al.} \cite{Kohsaka-nature-08}  found that the checkerboard pattern emerges most clearly in
Z-maps which plot the ratio of differential conductances $g({\boldsymbol{r}},V)$, at opposite bias, when the voltage
$V$ is chosen to be at the local value of the antinodal energy gap. In an earlier publication by the same group,
individual spectra $g({\boldsymbol{r}},V)$, where plotted for a series of tip positions placed above O-sites in a small
area with a pronounced checkerboard  modulation. \cite{Kohsaka-science-07}  This is illustrated in the checkerboard modulation
which shows up clearly in the R-map in Fig.\ref{fig:STM} which is reproduced from their Fig.4B. \cite{Kohsaka-science-07} The R-ratio measures the ratio
of the integrated currents $I({\boldsymbol{r}},V)$, at positive and negative voltages
\begin{eqnarray}
R({\boldsymbol{r}}, E=eV) &=& I({\boldsymbol{r}},+V) / I({\boldsymbol{r}},-V)     \label{eq:Rmap}
\end{eqnarray}
 The map reproduced in Fig. \ref{fig:STM} shows the R-pattern taken at an energy $E= 150$ meV, with clear
 modulations when the tip is above the O-sites with a local $C_{2}$ symmetry but almost no modulations are
 observed over the Cu-sites (denoted by black crosses). The value of the R-ratio over the O-sites ranges from 0.9 (shown as white) to 0.5 (black).

\begin{figure}[t]
\includegraphics[width = 8.5cm, height = 5.5cm, angle
= 0]{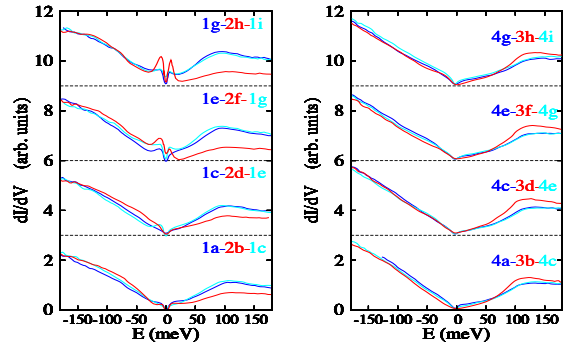} \caption[]{
(Color online) Differential tunneling conductance spectra taken on
oxygen sites along the vertical lines (1,2,3,4) as indicated in
Fig. \ref{fig:STM}, reproduced from Kohsaka \textit{et al.}
\cite{Kohsaka-nature-08}. Each subpanel shows the comparison of
$dI/dV$
 between the two neighboring O$_{x}$s (blue and cyan) and one O$_{y}$ (red) on each (Cu-O-Cu)$_{4}$ palette (with the position shown in Fig.\ref{fig:STM}).
 Note that the spectra is normalized to have the same intensity at negative voltages to facilitate the comparison and demonstrate the lowering of the
 symmetry from $C_{4}$ to $C_{2}$.}
\label{fig:STM_o}
\end{figure}

 The individual differential conductances taken above a series of O- and Cu- sites were also shown in their Fig. 5B.
 We have extracted the individual tunneling spectra $g({\boldsymbol{r}},V)$ above the O-sites in this typical small area with
 a pronounced checkerboard pattern in the R-map (shown in Fig. 5B of Ref[2]
 ). 
All spectra on these underdoped samples show a substantial asymmetry between
negative and positive 
voltages\cite{Anderson-jpcs-06,Randeria-prl-05}.  The negative voltage spectra,
which
correspond to electron extraction or hole injection,
show little structure and 
has the larger differential conductances. 
Since a hole can easily exchange positions
 with a neighboring occupied site, while an electron can only hop
 onto the unoccupied neighboring sites, the tunneling processes that accompany
an 
injected hole are much stronger than those for an electron, as shown by
Anderson and Ong~\cite{Anderson-jpcs-06}
and by Randeria \textit{et al}.  ~\cite{Randeria-prl-05}. Within the renormalized mean
field theory (RMFT) of
Zhang \textit{et al}. \cite{Zhang-88}, the asymmetry is largely due to the incoherent
tunneling process in 
the hole injection.  In the approach we use here based on the YRZ normal state
propagator, 
only coherent parts of the tunneling processes are included, and the
particle-hole asymmetry
appears at finite voltages which involves tunneling into quasi-particle states
away from the chemical potential.  The positive voltage spectra show a clear 
peak around an energy of order 100 meV. 
 Such a peak appears in the density of states calculated for the YRZ propagator by Yang \textit{et al.} \cite{Yang-prb-06}
 The YRZ propagator is an ansatz to describe coherent quasiparticles moving in a RVB background and so should agree better for positive voltage due to
 the weaker contribution of inelastic processes which accompany electron injection. In the subsequent discussion we will focus on the positive
 voltage spectra and normalize the spectra to be the same in the negative voltage region. In Fig. \ref{fig:STM_o} we show a series of the differential
 conductances with the tip above O-sites in the $\boldsymbol{x}/\boldsymbol{y}$ oriented  Cu-O-Cu bonds. The difference from the lower $C_{2}$-symmetry
  is clearly visible.

\begin{figure}[t]
\centerline{\includegraphics[width = 8.5cm, height = 6.5cm, angle
=0]{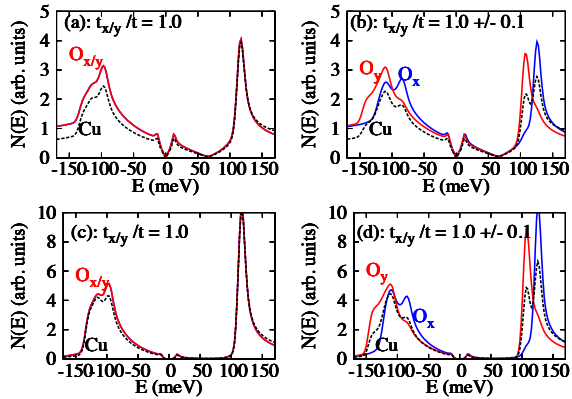}}
\caption[]{ (Color online) The density of states on Cu and two
kinds oxygen sites [O$_{\text{x/y}}$] from YRZ propagator
Eq.\ref{eq:DOS} without (panels a and b) and with (panels c and d)
the interference factor $(\cos{k_{x}} -\cos{k_{y}})^{2}$. Panels
(b, d) are the results in the presence of an intrinsic $C_{2}$ asymmetry in the NN hopping
integrals along $\boldsymbol{x}$ and $\boldsymbol{y}$ direction $t_{x/y}/t =
1\pm 0.1$. The splitting of the peaks at positive voltage shows up clearly. The other parameters are as following: $\Delta_{0} =
0.4$, $t^{\prime} = -0.3$, $t^{\prime\prime} = 0.2$, doping
$x=0.10$ (all energies are of unit $t = 300$meV). Very small
superconducting gaps are added to each band in pseudogap state, which does not change the RVB gap behavior (the two pronounced peaks at high energy $\sim$ 100meV) on O$_{\text{x/y}}$.
\cite{Kohsaka-science-07}
} \label{fig:DOS}
\end{figure}

\section{Possible Sources of the Anisotropy in the Tunneling Spectra.}
We now discuss a number of physical effects which could cause an anisotropy in the tunneling spectra.
\subsection{ Intrinsic $C_{2}$ asymmetry due to nematic or Pomeranchuk  order}
 An asymmetry between the two antinodal regions near ${\boldsymbol{k}}_{A,x} = (\pi,0)$ and ${\boldsymbol{k}}_{A,y} = (0,\pi)$ suggests
 that an intrinsic instability which breaks the $C_{4}$-symmetry  may be present in these samples. The possibility of such an instability has been discussed
 extensively in the literature. One set of these proposals focuses on a quantum electronic liquid crystal with nematic order.  \cite{Kivelson-nature-98, Kivelson-rmp-03,  Oganesyan-prb-01, Kao-prb-05}  A second on a Pomeranchuk instability which splits the two Van Hove singularities. \cite{Yamase-prb-05, Halboth-prl-00}
 Recent neutron scattering measurements of the low frequency spin fluctuations in YBCO samples have been
  interpreted as evidence for such intrinsic instabilities. \cite{Hinkov-science-08} In the presence of intrinsic asymmetry
  the NN hopping matrix elements in the $\boldsymbol{x}$- and $\boldsymbol{y}$- directions $t_{x}$ and $t_{y}$ will be slightly different.
  Assuming that the intrinsic asymmetry in the self energy in the YRZ propagator is small,  the change of NN hopping integrals $t_{x/y}$ leads to a relative displacement of
  the reference energy for the RVB splitting at ${\boldsymbol{k}}_{A,x}$ and ${\boldsymbol{k}}_{A,y}$. This in turn leads to a shift
  in the position of the energy gaps  $\Delta_{R}({\boldsymbol{k}}_{A},x)$, relative to the constant chemical potential. This is apparent in
  Fig. \ref{fig:DOS} which shows the predicted anisotropy arising from a uniform hopping asymmetry. 
  
  The density of states, $N(E)$, observed by putting
  the STM tip above O$_{x/y}$ and Cu sites takes the following form from Eq.(\ref{eq:Cucurrent})
\begin{eqnarray}
N^{O_{x/y}}(E) &\sim & \sum_{i,{\boldsymbol{k}}} Z^{i}_{{\boldsymbol{k}}} \cos^{2}{\frac{k_{x/y}}{2}} (\cos k_{x} -\cos k_{y})^{2}
 \delta{(E^{i}_{{\boldsymbol{k}}} -E)}  \notag \\
N^{Cu}(E) &\sim& \sum_{i, {\boldsymbol{k}}} Z^{i}_{{\boldsymbol{k}}}  (\cos k_{x} -\cos k_{y})^{2} \delta{(E^{i}_{{\boldsymbol{k}}} -E)}
\label{eq:DOS}
\end{eqnarray}
where $(\cos k_{x} - \cos k_{y})^{2}$ is the interference factor due to the hybridization with the hole states centered on the 4 NN Cu-sites as discussed earlier. 

  In a clean well ordered sample with intrinsic $C_{2}$ asymmetry order large domains should occur but in the highly disordered Ca$_{2-x}$Na$_{x}$CuO$_{2}$Cl$_{2}$
  samples there is a strong spatially varying electric field. The
linear coupling between the orientation of the domains and
the local quadrupole component of a random electric field acts as
a random orienting field on the order. Since the symmetry
of the order is Ising like, the random orienting field
will act similarly to a random field in an Ising model. It is well
known that the ground state of this Ising model consists of a
random array of domains, consistent with the orientational domains
observed in the checkerboard patterns. A word of caution, however,
is in order. The local anisotropy in the STM spectra does not seem
to be consistent with that predicted for this type  local asymmetry order. The
spectra in Fig. \ref{fig:STM_o}  show a $C_{2}$ -modulation pattern
arising from an overall reduction in the positive voltage spectra,
rather than a shift in the maxima that would signal the presence
of this intrinsic $C_{2}$ asymmetry in Fig.\ref{fig:DOS}.

The theoretical DOS for the STM tip located above Cu- and O-sites
shown in Fig.\ref{fig:DOS} display considerably sharper structure
in both cases, with and without the weighting factor $(\cos{k_{x}}
-\cos{k_{y})^{2}}$, than is evident in the experimental
conductances in Fig.\ref{fig:STM_o}. The theoretical DOS contain
two prominent features. One is the approximately linear drop from
negative to positive voltages leading to the minimum at $\simeq$
60meV. Interestingly this linear feature has been reported in
integrated photoemission studies on BSSCO samples by Hashimoto
\textit{et al}., \cite{Hashimoto-arxiv-08} as discussed recently
by Yang \textit{et al.}
 \cite{Yang-epl-09}. In the STM spectra there are signs of a minimum at positive voltages in the spectral 2h, 2f in Fig.\ref{fig:STM_o}, but at a smaller voltage and
 faint signs in the other spectra. The second feature is the very sharp peak at $\sim$ 100meV in the theoretical spectra and the much broader
 peaks at around the same energy in the STM spectra. For both discrepancies a possible remedy could be the much stronger disorder broadening in the
 local probe in the STM spectra as compared to spatially averaged spectra observed in photoemission.

\subsection{ Random electric fields}
  The random distribution of Na$^{+}$-acceptors in Ca$_{2-x}$Na$_{x}$CuO$_{2}$Cl$_{2}$ samples generates a
  random electric field. The local quadrupole component of this field
acting on a (CuO)$_{4}$ plaquette will couple differently to the
states at ${\boldsymbol{k}}_{A,x}$ and ${\boldsymbol{k}}_{A,y}$. The
former is anti-bonding on the Cu-O-Cu bonds along the $\boldsymbol{x}$-axis and
bonding on the Cu-O-Cu bonds along the $\boldsymbol{y}$-axis while the latter
has the opposite bonding pattern. These patterns cause opposite
quadrupole charge distributions in the two states leading to a
potential splitting of the antinodal states in the presence of an external
electric quadrupole field. This potential splitting will generate STM spectra
similar to those we found for hopping asymmetry and so it also differs from
the anisotropy displayed in Fig. \ref{fig:STM_o}

\subsection{ Variations in the quasiparticle weight and in the localization length}
The tunneling DOS into quasiparticle states is weighted by the
quasiparticle weight in the single particle propagator. In the YRZ
ansatz this weight is taken from the RMFT calculations of Zhang
\textit{et al.} \cite{Zhang-88} and is simply proportional to the
hole density, $x$.  However the fact that little change in the
energy gaps $\Delta_{R}({\boldsymbol{k}}_{A,x},x)$, is evident in
the spectra shown in Fig. \ref{fig:STM_o} argues against such an
explanation. 
Also the spatial variation in the hole density is longer ranged
than the distance between O-sites on a (CuO)$_{4}$. Another
possibility is that there is substantial anisotropy in the
disorder scattering in ${\boldsymbol{k}}$-space which causes a
local anisotropy in the localization lengths. An anisotropic
energy shift and localization length could result, if the potential scattering from
the disorder is mainly  between opposite antinodal points rather than between
$ \pm \boldsymbol{k}_{A,x}$ and $\pm \boldsymbol{k}_{A,y}$. An anisotropic energy shift, for example, could remove weight from the energy range (~200meV) covered in the STM experiments. The somewhat longer range of the disorder potential is
consistent with this suggestion. The umklapp correlated scattering
wave vector for backscattering between the antinodal regions near
$\pm \boldsymbol{k}_{A,x}$ (or $\pm \boldsymbol{k}_{A,y}$) is much
smaller than the wave vector connecting the two antinodal regions,
$\simeq (\pi, \pi)$. The Fourier transforms of the checkerboard
maps show peaks arising from short range order in the form of $4a$
wide unidirectional electronic domains. The analysis presented
here does not directly address the presence of short range order
on a length scale $>a$. Several authors have suggested that the
wave vector of this short range order is approximately equal to
the wave vector connecting parallel pieces of the bare band
structure Fermi surface near the saddle points \cite{Wang-prb-03}
and also the wave vector connecting the turning points of the
quasiparticle contours in the normal state YRZ propagators.
\cite{leni-prb-08} The length scale of the domains at $\sim 4a$ is
roughly the same as the disorder potential length scale but the
interpretation of the short range order in the checkerboard
patterns remains to be an open issue.

\section{Conclusion} In this paper we have examined several
possible origins for the local checkerboard pattern in the
tunneling DOS measured over O$_{\text{x/y}}$-sites. The presence of interference
between two tunneling paths centered above the neighboring
Cu-sites leads to an anisotropic weighting of the two antinodal
regions in ${\boldsymbol{k}}$-space. Thus the local $C_{2}$ pattern
of the checkerboard can reflect an asymmetry between the two
antinodal ${\boldsymbol{k}}$-space regions. 
An asymmetric hopping ($t_{x}$, $t_{y}$) could arise from an intrinsic instability e.g. towards nematic or Pomeranchuk order. This would split the energies of the pseudogaps at $\pm \boldsymbol{k}_{A,x}$ and $\pm \boldsymbol{k}_{A, y}$.
However, an examination of the measured
anisotropy in the differential conductance points rather towards
an asymmetry in the average magnitude of the conductance, which possibly could result from a difference in the localization lengths at the two antinodal
saddle points in the quasiparticle dispersion.

We are very grateful to J.C.S. Davis and Y. Kohsaka for permission
to reproduce their STM data  and Y. Chen for
stimulating discussions. Support from the MANEP program of the
Swiss National funds (K.-Y. Y. and T. M. R.), and RGC grant of
HKSAR (W.-Q. Chen and F. C. Z.) is gratefully acknowledged. 
This research was supported in part by the National Science Foundation under Grant No. PHY05-51164.

\end{document}